\documentclass[a4paper,twoside]{article}
\usepackage{AMS2LA,fancyhdr,CJK,multicol,graphics,bm}
\usepackage[dvipdfm, a4paper, pdfstartview=FitH, bookmarks=false, colorlinks=false,
        linkcolor=blue, urlcolor=blue, pdfborder=001, citecolor=blue,
        pdftitle={Changepdftitle}, pdfauthor={Changepdfauthor},
        pdfcreator=CHIN.\ PHYS.\ LETT., pdfproducer=CPL,
        pdfkeywords={changePACS}]{hyperref}
\def\headrule{\kern 1mm \hrule width 17cm \kern -1mm}%
\def\footnoterule{\kern 1mm \hrule width 7cm \kern 2.2mm}%
\def\REF#1{\par\hangindent\parindent\indent\llap{#1\enspace}\ignorespaces}%
\renewcommand{\arraystretch}{1.1}     % table-distance
\newcommand{\cplyear}{2013} \newcommand{\cplvol}{30}
\newcommand{\cplno}{1} \newcommand{\cplpagenumber}{01{2101}}
 \newcommand{\cplpage}{\cplpagenumber-\thepage}
\begin{document} \begin{CJK}{GBK}{song}\vspace* {-6mm} \begin{center}
%---------------------------Title --------------------------------
\large\bf{\boldmath{To differentiate neutron star models by X-ray polarimetry}}
%--------------------------Footnote--------------------------
\footnote{This work is supported by the National Basic Research
Program of China (2012CB821800, 2009CB824800) and the National
Natural Science Foundation of China (11225314, 10935001, 10973002).

\hspace*{1.8mm}$^{\S}$Correspondence author. Email: {\tt
r.x.xu@pku.edu.cn}

\hspace*{1.8mm}\copyright\,{\cplyear}
\href{http://www.cps-net.org.cn}{Chinese Physical Society} and
\href{http://www.iop.org}{IOP Publishing Ltd}}
\\[4mm]
%---------------------------Authors-----------------------------
\normalsize \rm{}Lu Ji-Guang$^{1}$, Xu Ren-Xin$^{1\S}$, Feng
Hua$^{2}$
%--------------------------COM. or University -------------------
\\[1mm]\small\sl $^{1}$School of Physics and State Key Laboratory of Nuclear Physics and Technology, Peking University, Beijing 100871

$^{2}$Department of Engineering Physics and Center for Astrophysics,
Tsinghua University, Beijing 100084
%------------------------ Received date----------------------
\\[4mm]\normalsize\rm{}(Received ...)
\end{center}
\end{CJK}
%----------------------Abstract and PACS---------------------
\vskip -1mm

\noindent{\narrower\small\sl{}The nature of pulsar is still unknown
because of non-perturbative effects of the fundamental strong
interaction, and different models of pulsar inner structures are
then suggested, either conventional neutron stars or quark stars.
Additionally, a state of quark-cluster matter is conjectured for
cold matter at supranuclear density, as a result pulsars could thus
be quark-cluster stars. Besides understanding different
manifestations, the most important issue is to find an effective way
to observationally differentiate those models. X-ray polarimetry
would play an important role here. In this letter, we focus on the
thermal X-ray polarization of quark/quark-cluster stars. While the
thermal X-ray linear polarization percentage is typically higher
than $\sim 10\%$ in normal neutron star models, the percentage of
quark/quark-cluster stars is almost zero. It could then be an
effective method to identify quark/quark-cluster stars by soft X-ray
polarimetry. We are therefore expecting to detect thermal X-ray
polarization in the coming decades.

\par}\vskip 3mm\normalsize

\noindent{\narrower\sl{PACS: 95.55.Qf, 97.60.Gb, 26.60.Kp, 95.55.Ka, 95.75.Hi}%
{\rm\hspace*{13mm}DOI:???/\cplvol/\cplno/\cplpagenumber}

\par}\vskip 3mm
%--------------------TEXT TEXT TEXT TEXT---------------------
\begin{multicols}{2}

Pulsar study is not only important in understanding diverse
phenomena of high-energy astrophysics, but also significant in
fundamental physics.
The nature of the compressed baryonic matter in pulsars is still not
certain because of the non-perturbative effects of the fundamental
color interaction.$^{[1]}$
In view of the high density, there are two types of models:
gravitation-bound or self-bound ones. Normal neutron star model is a
typical representative of the former, while quark/quark-cluster
model belongs to the latter.
Although both of the two models might explain the thermal X-ray
spectra of pulsars, the polarization behaviors would be quite different.

A normal neutron star (more generally, hadron star or mixed star) as
gravitationally confined object must have an atomospheric envelope
composed by normal matter with pressure gradient to link high
pressure interior and the zero pressure outside, but this envelope
would not be necessary for self-bound body, such as bare quark star
or quark-cluster star.
Phenomenologically, some observations may hint that a bare and
self-confined surface might exist in order to naturally understand
different observational manifestations (e.g., sub-pulse drifting,
non-atomic spectrum, clean fireballs for supernova/$\gamma$-ray
burst).$^{[2]}$
%
%It was expected that the composition of quark-cluster star's surface
%is also degenerate matter. In fact there are a lot of evidence showing
%the surface of pulsars is strongly self-confined.$^{[2]}$
%
It was expected that, because of low temperature gradient of surface
with degenerate electrons, the linear polarization of thermal X-ray
emission from quark-cluster star would be very low,$^{[3]}$ however
a quantitative calculation has never been presented.
In this paper, we calculate the polarization behavior of
quark-cluster star and compare our result to pre-existing conclusion
of neutron star $^{[4]}$ in order to test pulsar structure models by
future advanced X-ray polarimetries.

There are two mechanisms for generating thermal X-ray polarization
of pulsar. The separatrix is the critical magnetic field, $B_q\simeq
4\times10^{13}$ G. For weak magnetic field, i.e. $B<B_q$, quantum
vacuum effect could be negligible.

When X-rays propagate across magnetic B-field, there are two
independent linear polarization eigenmodes: ordinary mode (O-mode,
electric field in the plane of wave vector and B-field) and
extraordinary mode (E-mode, perpendicular to the plane), but the
opacity coefficients of a magnetized thermal plasma are different
for them.
Gnedin \& Sunyaev (1974) presented an approximation about the cross
section of photon-electron scattering for photon frequency
$\omega<<\omega_c\equiv eB/m_ec=11.6\times B_{12}~\mathrm{keV}$
($m_e$ is the mass of electron and $B_{12}=B/10^{12}$ G) and angle
between the wave vector and the B-field
$\theta>(\omega/\omega_c)^{1/2}$,$^{[5]}$
$$
\sigma_O=\sigma_T \sin^2 \theta,  \tag 1
$$
$$
\sigma_E=\sigma_T (\omega/\omega_c)^2 (1/\sin^2 \theta), \tag 2
$$
where $\sigma_O$, $\sigma_E$ is the cross sections of O-mode and
E-mode, respectively, and $\sigma_T$ is the Thomson scattering cross
section. For normal pulsars of $B\simeq 10^{12}$ G,
$\theta\gg(\omega/\omega_c)^{1/2}$, one has $\sigma_O\gg\sigma_E$.
This implies that the average free path length of O-mode photon
$L_1$ is far less than that of E-mode photon $L_2$ (see Fig.\,1,
i.e., different photospheres for those two modes), and hence the
optical depth depends on its polarization behavior. Due to
temperature gradient, E-mode intensity would be much higher than the
O-mode one, and the thermal X-rays are thus polarized.
Therefore, X-ray polarimetry would thus provide a measurement of
pulsar surface temperature gradient.
Pavlov \& Zavlin (2000) had concluded that the linear polarization
of normal neutron star could be as high as 10\%-30\%.$^{[4]}$

In case of magnetic field $B>B_q$, additional quantum vacuum effect
due to quantum electrodynamics (QED) will also cause polarization of
thermal X-ray radiation.$^{[6-8]}$ Lai \& Ho$^{[8]}$ demonstrated
that a QED vacuum effect called vacuum birefringence emerges for
$B\geq7\times10^{13} \mathrm{G}$, and found a very high average
polarization at 10\%-100\% for magnetars.$^{[7]}$

\vskip 0.5\baselineskip

\vskip 4mm

\centerline{\includegraphics{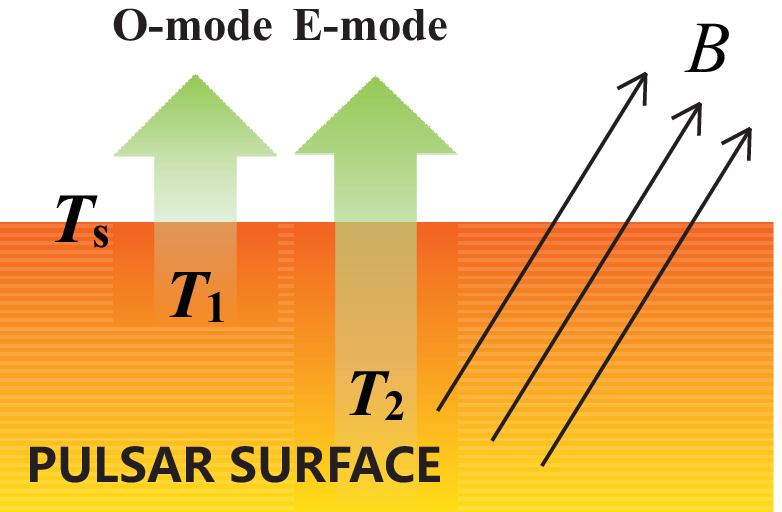}}

\vskip 2mm

\centerline{\footnotesize \begin{tabular}{p{7.5 cm}}\bf Fig.\,1. \rm
A schematic diagram of thermal X-ray polarization originated from
pulsar surface, where the QED vacuum polarization effects are not
included. The E-mode photons come from deeper and hotter place than
that of the O-mode.
\end{tabular}}

\medskip

Above are previous results of thermal X-ray polarization of normal
neutron stars/magnetars. For comparison, we are calculating the
thermal X-ray polarization in quark-cluster star model, as
following.

Thermal conductivities ($\kappa$) of degenerate electrons inside
quark or quark-cluster stars can be conveniently expressed through
effective electron collision frequencies, $\nu_{ee}$,$^{[9]}$
$$
\kappa=\frac{\pi^2k_B^2T_Sn_e}{3m_e\nu_{ee}},  \tag 3
$$
where $n_e$, $T_S$ denote the number density of the electron and the
temperature of the quark star surface, respectively, and $k_B$ is
the Boltzmann constant. The effective electron collision frequencies
can be derived by following formula,$^{[10]}$
$$
\nu_{ee}\simeq\frac{3}{2\pi}(\frac{\alpha}{\pi})^{1/2}\frac{(k_BT_S)^2}{\hbar\varepsilon_F}J(\varsigma),  \tag 4
$$
$$
J(\varsigma)=\frac{1}{3}\frac{\varsigma^3ln(1+2\varsigma^{-1})}{(1+0.074\varsigma)^3}+\frac{\pi^5}{6}\frac{\varsigma^4}{(13.9+\varsigma)^4},  \tag 5
$$
$$
\varsigma=2\sqrt{\frac{\alpha}{\pi}}\frac{\varepsilon_F}{k_BT_S},  \tag 6
$$
where $\alpha=e^2/\hbar c$ is the fine structure constant, and
$\varepsilon_F=\hbar c(\pi^2n_e)^{1/3}$ is the Fermi energy of
degenerate electrons.

In order to explicate that the polarization of thermal radiation
from quark-cluster star is small enough to be ignored, we calculate
the maximum linear polarization ($P_{\rm max}$) just for $\theta=90^{\circ}$,
$$
P_{\rm max}=\frac{|J_O-J_E|}{J_O+J_E}\sim \frac{|\sigma T_1^4-\sigma
T_2^4|}{\sigma T_1^4+\sigma
T_2^4}=\frac{|T_1^4-T_2^4|}{T_1^4+T_2^4},  \tag 7
$$
where $J_O$ and $J_E$ is the X-ray intensity of O-mode and E-mode,
respectively, $T_1$ ($T_2$) is the average temperature where the
O-mode (E-mode) photons could come out from, and $\sigma$ is the
Stefan-Boltzmann constant.
The thermal conductivities of strange quark-cluster matter is
extremely high, the temperature gradient would then be very small.
Therefore, approximation $T_S-T_1\ll~T_S-T_2\ll~T_S$, and Eq.\,(7)
will be reasonable,
$$
P_{\rm
max}\simeq\frac{|T_S^4-T_2^4|}{T_S^4+T_2^4}\simeq\frac{T_S^3\cdot\Delta
T}{2T_S^4}=\frac{\Delta T}{T_S}, \tag 8
$$
where $\Delta T\equiv  T_2-T_S$.

For the approximation of black body radiation, the energy flux
density $J_r$ is,
$$
J_r=\sigma T^4,  \tag 9
$$
but for thermal conduction, the energy flux density $J_c$ is
expressed as,
$$
J_c=\kappa \cdot \bigtriangledown T\simeq\kappa\frac{\Delta T}{L_2}.
\tag 10
$$
One has $J_r=J_c$ since there is no energy source near quark-cluster
star surface. Combining equation Eq.\,(9) and Eq.\,(10),
$$
\Delta T=\frac{\sigma~T_S^4L_2}{\kappa}. \tag 11
$$
Considering the propagation of E-mode photons, we could have the
free path length,
$$
L_2\simeq \frac{1}{n_e\sigma_E}\frac{\varepsilon_F}{k_BT_S}, \tag 12
$$
where a factor of $\varepsilon_F/k_BT_S$ is introduced because only
electrons near the Fermi surface could scatter off the X-rays.

According to the equations of Eq.\,(2), Eq.\,(3), Eq.\,(8),
Eq.\,(11) and Eq.\,(12), one comes to,
$$
P_{\rm
max}\simeq\frac{6\sigma~T_Sm_e\nu_{ee}\omega_c^2\varepsilon_F}{\pi^2k_B^3n_e^2\sigma_{T,corr}\omega^2},
\tag 13
$$
i.e., $P_{\rm max}\propto\omega^{-2}$, where the relativity
correction is included.$^{[11]}$.

We calculate the maximum linear polarization (to maximize the
polarization, we just consider the head-on collisions of photons and
electrons) for typical parameters of $n_b=1.5 n_0$ and $n_e=10^{-4}
n_b$, where $n_b$ is the number density of baryon in quark-cluster
star, with $n_0$ the number density of nuclear matter. The results
are shown in Fig.\,2, which shows that the polarization of thermal
radiation from a quark-cluster star is too small to detect.

\vskip 0.5\baselineskip

\vskip 4mm

\centerline{\includegraphics{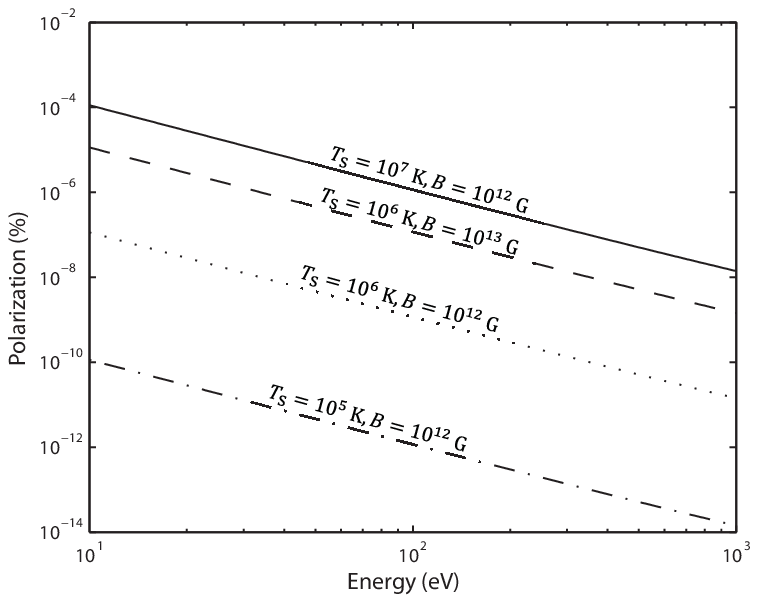}}

\vskip 2mm

\centerline{\footnotesize \begin{tabular}{p{7.5 cm}}\bf Fig.\,2. \rm
The thermal X-ray polarization of quark-cluster star as a function
of photon energy, for parameterized temperatures ($T$) and magnetic
fields ($B$) illustrated.
\end{tabular}}

\medskip

There is an unexceptionable source to test the models, RX
J1856.5-3754. Discovered in 1996,$^{[12]}$ it is the brightest one
in all the isolated neutron stars.
The X-ray spectrum of RX~J1856.5$-$3754 can be adequately fitted
by a blackbody spectrum.
The non-variable thermal spectrum show that we do indeed see the surface
of this pulsar directly.
It is always controversial about the state of matter for a very
stiff equation of state (EoS) constrained by its small
radius,$^{[13,14]}$ although the stiff EoS could be naturally
understood by a Lennard-Johns quark matter model.$^{[15]}$
The neutron star model needs a very strong magnetic field to explain
the absence of spectral lines, while the quark/quark-cluster star
model doesn't need.$^{[16]}$

In the regime of normal neutron star, the featureless Planckian
spectrum of RX J1856.5-3754 may hint a superstrong B-field, in which
unique signatures of the vacuum polarization emerge. The field would
be so strong that the outermost layer might be in a condensed solid
or liquid.
We can also calculate the polarization of the neutron star in the
model provided in Ref.[17], and the results are shown in Fig.\,3,
with the photon energy to be fixed at 0.25 keV.
It is evident that significant linear polarization could also be
detectable even if the B-filed is really so strong that the surface
is condensed.
It is worth noting that the observed X-ray flux peaks at a few
hundreds electron-volts, where X-ray polarization can be measured
using the multilayer based polarimeter.$^{[18]}$

\centerline{\includegraphics{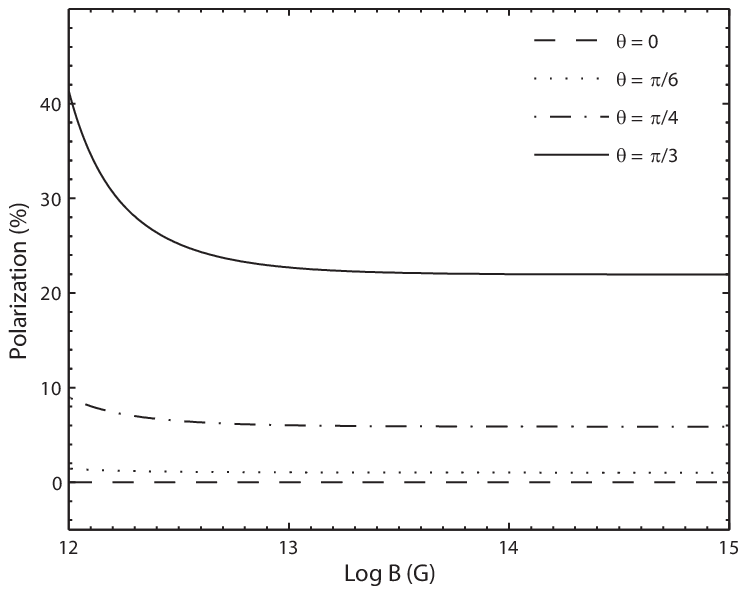}}

\vskip 2mm

\centerline{\footnotesize \begin{tabular}{p{7.5 cm}}\bf Fig.\,3. \rm
The X-ray polarization of thermal radiation directly from the
degenerate metallic condensed surface with strong magnetic field
($B$). The vacuum polarization of QED is included, for photon energy
at 0.25 keV but different emergence angles ($\theta$, the angle
between magnetic field and the wave vector), as illustrated.
\end{tabular}}

\vskip 2mm

Soft $\gamma$ repeaters (SGRs) and anomalous X-ray pulsars (AXPs)
are all magnetar candidates. However, it is not necessary to assume
such a strong field to explain the large period derivative and
enormous energy release in the solid quark-cluster star
model.$^{[19]}$
Nonetheless, energy release due to magnetic field reconnection would
still be significant in order to understand the observations of
SGR/AXPs (especially that of the superflares) in conventional liquid
quark star models (e.g., in a magnetic CFL phase $^{[20]}$).
Therefore, X-ray polarimetry could also be a powerful way to test
the magnetar model.

In summary, we have shown that X-ray polarimetry will be a powerful
tool to differentiate neutron star models. For normal neutron
star/magnetar models, the linear polarization of thermal X-rays
would be high enough to be detectable. On the contrary, the
polarization of thermal radiation from quark-cluster stars would be
truly negligible.
The brightest compact object RX J1856.5-3754, with pure thermal
radiation, should be an idea source for the soft X-ray polarization
observation and a testbed of compressed baryonic matter problem.
It is really worth verifying the conjectures by advanced X-ray
polarimetry.

The distinct thermal polarization predicted for normal neutron stars
and quark/quark-cluster stars can be readily tested with future soft
X-ray polarimeters, for example, the Lightweight Asymmetry and
Magnetism Probe (LAMP) project being developed in China. LAMP will
detect X-ray polarization at 250 eV using multilayer mirrors at
incidence angles near 45 degrees with a sensitivity, in terms of
minimum detectable polarization, of 5\% or less for objects as
bright as RX J1856.5-3754. Therefore, it is capable of
distinguishing those two competing models.

\vspace{2mm}
\noindent %
{\bf Acknowledgments}
We would like to thank Prof. Shuangnan Zhang to encourage our work
of this paper and acknowledge useful discussions at the pulsar group
of Peking University.

\section*{\Large\bf References}%

\vspace*{-0.8\baselineskip}

\hskip 7pt {\footnotesize%

\REF{[1]} Xu R X 2011 {\it Int. Jour. Mod. Phys. E} {\bf 20} 149

\REF{[2]} Dai S and Xu R X 2012 {\it International Journal of Modern Physics: Conference Series} {\bf 10} 137

\REF{[3]} Xu R X 2003 {\it Astrophys. J.} {\bf 596} L59

\REF{[4]} Pavlov G G and Zavlin V E 2000 {\it Astrophys. J.} {\bf 529} 1011

\REF{[5]} Gnedin Yu N and Sunyaev R A 1974 {\it Astron. and Astrophys.} {\bf 36} 379

\REF{[6]} Heyl Jeremy S and Shaviv Nir J 2002 {\it Physical Review D} {\bf 66} 3002

\REF{[7]} Lai Dong and Ho Wynn C G 2003 {\it Physical Review Letters} {\bf 91} 1101

\REF{[8]} Lai Dong and Ho Wynn C G 2003 {\it Astrophys. J.} {\bf 588} 962

\REF{[9]} Urpin V A and Yakovlev D G 1980 {\it Soviet Astronomy} {\bf 24} 425

\REF{[10]} Usov Vladimir V 2001 {\it Astrophys. J.} {\bf 550} 179

\REF{[11]} Heilter W 1954 {\it The Quantum Theory of Radiation} (Oxford, London)

\REF{[12]} Walter Frederick M, Wolk Scott J and Neuh$\mathrm{\ddot{a}}$user 1996 {\it Nature} {\bf 379} 233

\REF{[13]} Braje Timothy M and Romani Roger W 2002 {\it Astrophys. J.} {\bf 580} 1043

\REF{[14]} Walter Frederick M 2004 {\it Journal of Physics G: Nuclear and Particle Physics} {\bf 30} 461

\REF{[15]} Lai X Y and Xu R X 2009 {\it Mon. Not. R. Astron. Soc.: Lett.} {\bf 398} 31

\REF{[16]} Xu R X 2002 {\it Astrophys. J.} {\bf 570} L65

\REF{[17]} van Adelsberg M et al 2005 {\it Astrophys. J.} {\bf 628}
902

\REF{[18]} Marshall H L et al 2003 {\it Astronomical Journal} {\bf
125} 459

\REF{[19]} Tong H and Xu R X 2011 {\it International Journal of Modern Physics E} {\bf 20} 15

\REF{[20]} Ferrer E J and de la Incera V 2013  in {\it Lecture Notes
in Physics}, ``Strongly interacting matter in magnetic fields''
(e-print arXiv:1208.5179)

 }
\end{multicols}
\end{document}